\documentclass[11pt, a4paper]{revtex4-2}

\usepackage[a4paper, total={18cm, 24.5cm}]{geometry}
\usepackage{graphicx}
\usepackage{setspace}
\usepackage{amsfonts}
\usepackage{amsmath}
\usepackage{pxfonts}
\usepackage{enumitem}
\usepackage{datetime}

\begin{document}
\begin{singlespace}
\title{Conservation in High-field Quantum Transport}

\author{Mukunda P. Das$^1$ and
Frederick Green$^{2,}$
}
\email{frederickgreen@optusnet.com.au}
\affiliation{
$^1$ 
Department of Fundamental and Theoretical Physics,
RSP, The Australian National University, Canberra, ACT 2601, Australia.\\
$^2$ 
School of Physics, The University of New South Wales,
Sydney, NSW 2052, Australia.
}


\begin{abstract}
We give a short overview of the role of microscopic conservation in charge
transport at small scales, and at driving fields beyond the linear-response
limit. As a practical example we recall the measurement and theory of
interband coupling effects in a quantum point contact driven far from
equilibrium.
\end{abstract}
\maketitle


\section{Introduction}

It is a given that any description of charge motion in an extended
system must satisfy the continuity Equation
\cite{NP}
\begin{eqnarray}
\frac{\partial \rho}{\partial t}
&=&
-\nabla\cdot{\bf j}
\label{01}
\end{eqnarray}
tying the loss(gain) of local charge carrier {\em number} density
$\rho({\bf r},t)$, as a function of position and time, to the
outflow(inflow) of the local carrier flux ${\bf j}({\bf r},t)$.
This dynamical identity is always accompanied by the static,
or at least slowly time varying, Poisson Equation
\cite{NP}
\begin{eqnarray}
\nabla\cdot{\bf E} = - {\nabla^2} \phi({\bf r}, t)
&=&
\frac{e}{4\pi\varepsilon_0\varepsilon}
[\rho_0({\bf r}) - \rho({\bf r}, t; \phi)]
\label{02}
\end{eqnarray}
where $\phi({\bf r}, t)$ is the local, conservative, electric potential and
$\rho_0({\bf r})$ is the compensating background ion-charge number density
given at equilibrium, with $\varepsilon$ the relative permittivity of the
medium. Here we take $\rho$ to describe the number density of mobile
electrons.

Equations (\ref{01}) and (\ref{02}) define an electrical transport problem,
be it classical or quantum
\cite{fren},
in which two exact conservation laws describe
three coupled unknown functions. The practical issue is, naturally, to
establish additional relations among them to close the computation. This
is impossible to achieve exactly, including realistic situations such
as quantum transport in conductors at the meso- and nanoscopic scales.
A relevant model is the classic inverse Ohm's law, which posits
${\bf j} \equiv {\boldsymbol \sigma}\cdot{\bf E}$ with
${\boldsymbol \sigma}$ the
phenomenological conductivity tensor.

In this abridged review we first recall the elements underlying any
microscopic description of circuit transport. In the second part we analyse a
specific example: high-field current response in a quantum point contact
(QPC), in which interband transitions are prominent. A minimum set of
assumptions leads to a practicable transport solution that builds in
conservation from the start.
Their detailed outworking for the interband problem is in Reference
\cite{GD2}.
A brief summary ends this paper.

\section{Physical Picture}

\begin{figure}
\centerline{
\includegraphics[height=8.0 cm]{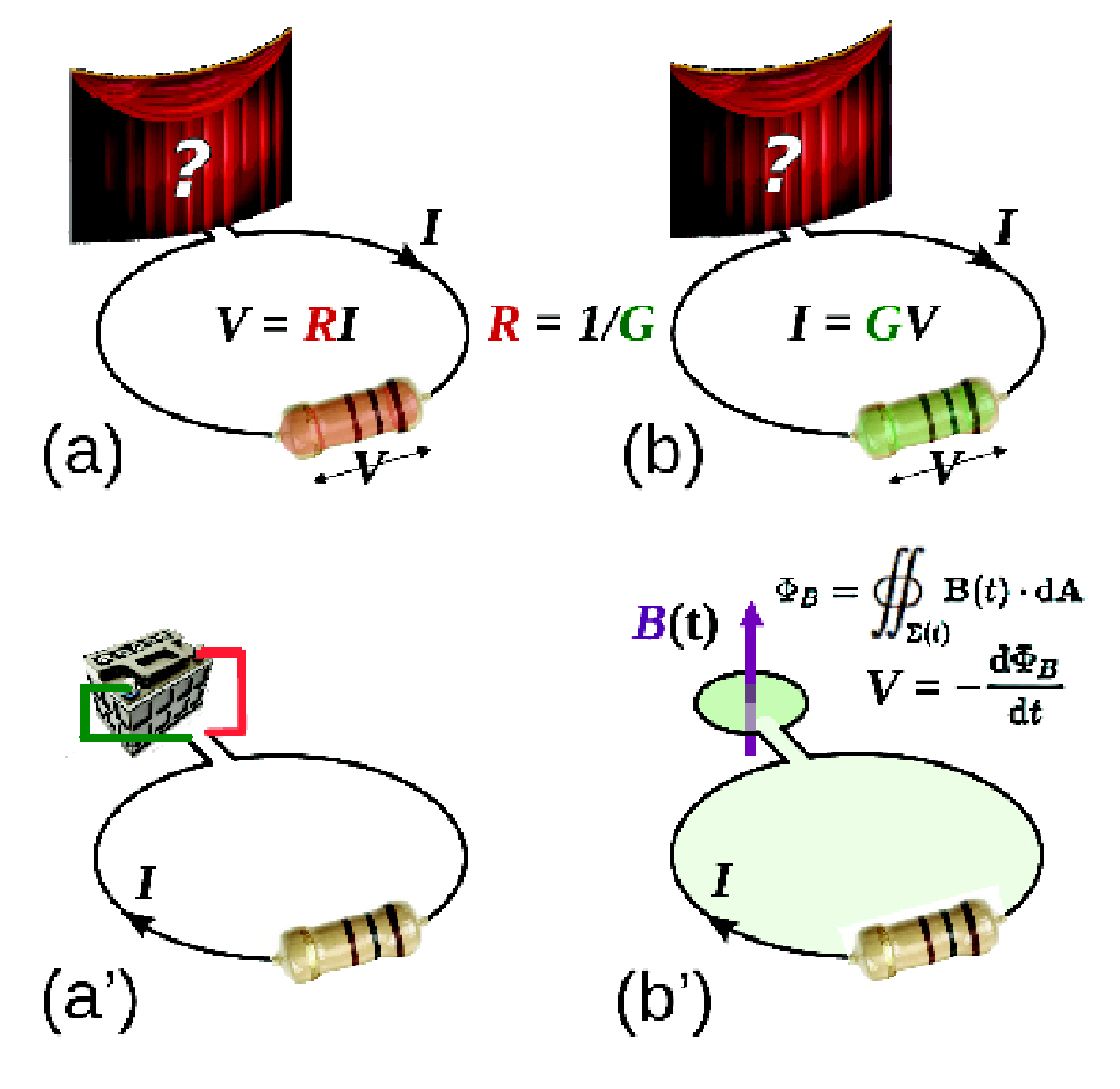}
}
\caption{Operational equivalence of a battery and a time-varying magnetic
flux as nonconservative current sources for a generic closed circuit.
As power supplies, both have lifetimes that are finite but sufficiently long
relative to the time scale for measurement. With both sources notionally
concealed, as in (a) and (b), there is no means to discriminate between the
battery, (a'), and the changing flux, (b'), using only measurements on
the active device. In either case the electromotive force cannot be described
by a conservative potential. Such a potential is incapable of sustaining a
constant circulating current. 
\cite{WW,kk}.
\label{F1}}
\end{figure}

The fundamental premise underlying the entire study of transport is
electrical neutrality of the whole system
as is, of course, true
\cite{note1}
for every active electrical circuit. Over a system volume $\Omega$
this entails the {\em unconditional} constraint
\cite{NP}
\begin{equation}
\int_{\Omega} d{\bf r} [\rho({\bf r}, t;\phi) - \rho_0({\bf r})]
\equiv
0
= \oiint_{\Sigma(\Omega)} \nabla \phi \cdot  d{\bf A}
\label{03}
\end{equation}
The rightmost integral over the system surface $\Sigma(\Omega)$ expresses
Gauss' theorem implying, from Equations (\ref{02}) and (\ref{03}), that
any field line ${\bf E} = -\nabla\phi$ entering the neutral system must
also leave it
\cite{note3}.
The potential generating such a field must be conservative and could not
in itself sustain a net current across the bulk structure.

This begs the question of how to drive a current in steady state. The answer
lies in the fact that a stable current (one possessing a period
much longer than any measurement time) is physically possible
only in a circuit geometry that is
\begin{enumerate}
\item{}
multiply connected globally;
\item{}
electrically neutral overall;
\item{}
driven by a {\em nonconservative} internal field
(the electromotive force, or EMF);
\item{}
admits a mechanism dissipating absorbed power to the environment. 
\end{enumerate}

The paradigm of a nonconservative driver is a battery, but a time-varying
magnetic flux threading the circuit loop is conceptually ideal as a driver
(and as applied in power engineering), manifesting Faraday's law
\cite{WW}.
In Figure \ref{F1} we illustrate the operational equivalence of the
two nonconservative ways of driving a conducting device. 

In setting up a conserving transport description, the next step is to
analyse the role of the macroscopic leads injecting and extracting the
globally circulating current where it transits the active region.
By their screening action, the strongly metallic leads re-establish total
electrical neutrality within a very short distance (nanometres and less)
from the regions where local accumulation and depletion of charge occurs
at the interfaces with the lithographically defined device structure, which
in the following will be a QPC.

Metallic screening at the boundaries means that the neutrality condition,
Item 2 in the preceding list, applies more strictly to the QPC structure
but not only that alone. It must be treated {\em together with} the
immediate interfaces to the leads connecting it to the driving
source. The interfaces are of one piece with the nominal device structure.
This means that a proper description will cover the device
region up to and including the screening layers.

Charge neutrality does not depend on the EMF or the current.
{\em The electrochemical potential in the bulk leads always remains fixed}
\cite{kk},
Passive mismatch of conservative electrochemical potentials is
not the correct physical boundary condition for sustaining steady state
\cite{WW}.

A convenient way to conceive (not necessarily to implement) how to
include the preceding physics into a theoretical description is to
revise the continuity Equation for number density and current
when driven by an external pump supplying current $I$ to the active
region; see also Figure \ref{F2}.
Simplifying to one dimension, Equation (\ref{01}) is extended by
introducing a point source and drain of particles in the right-hand side:
\begin{eqnarray}
\frac{\partial \rho(x,t)}{\partial t}
+ \frac{\partial J(x,t)}{\partial x}
&=&
I{\Bigl( \delta(x + x_0) - \delta(x - x_0) \Bigr)}.
\label{04}
\end{eqnarray}

In Equation (\ref{04}) current is injected at location $-x_0$ standing in for
the ``upstream'' boundary and reabsorbed at $x_0$ standing in for the
``downstream'' boundary. The operational length of the active region is
then $L = 2x_0$. It is identifiable with the dominant mean free path (MFP)
for scattering, since this is determined by collisions with the lead
material at the interface regions. A general analysis of the electrodynamics
of this scenario, from a fully microscopic viewpoint, was given by Sols
\cite{sols}
who analysed the boundary conditions for the quantum Kubo formula
\cite{kk}.
A formally different but equivalent microscopic description is found
in Magnus and Schoenmaker
\cite{WW}.

To complete the formal framework one still needs two ingredients:
(a) a functional connection between the local current distribution $J$
and the local density $\rho$, and (b) a dissipative model of how the
device material sheds the electrical energy pumped into the system's
free energy by the power supply.
Both of these can be accommodated by extending the form of the
continuity Equation (\ref{01}), and its generalisation Equation (\ref{04}),
to describe the carrier's momentum distribution as it evolves. This
requires the kinetic quantum Boltzmann formalism, more fully elaborated
in Ref.
\cite{GD3}.

\section{A Nonequilibrium Problem}

\subsection{Setting the scene}

\begin{figure}
\centerline{
\hskip 5mm
\includegraphics[height=7.0 cm]{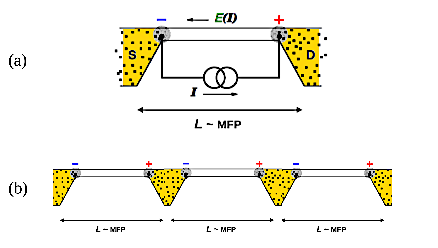}
}
\vskip 5mm
\caption{(a) Schematic of a ballistic quantum point contact driven by
an external current source and stabilised by bulk resistive metallic leads.
Strong metallic screening at the interfaces enforces electrical neutrality
in and charge uniformity over most of the device, independent of current
$I$ and EMF $E(I)$. The scattering mean free paths, both elastic and
dissipative, are bounded by the length scale including not only the
impurity-free QPC geometry but also the adjacent regions of strong boundary
scattering, where current enters and exits. The three structures together
make up the complete physical device, whose operational length $L$
delimits the longest mean free path. (b) An ensemble of identical such
structures in series is concpetualised and described on the average. The
current source and sink are now removed to the extremes of the ensemble.
}
\label{F2}
\end{figure}

In this Section we discuss the QPC transport experiment of
de Picciotto {\em et al}
\cite{1,dp2}.
This experiment graphically reveals the need to go beyond
single-carrier linear response to understand carrier kinetics
at mesoscopic scales and below.

Our calculations to support such an understanding were done within the
quantum Boltzmann protocol.
In brief, the microscopic analogue to Equation (\ref{01}) extends the
object of interest to the distribution of electron number density
$f_k(x,t)$ not only locally but in the space of available momentum
states $k$, where the external field acts to promote
them to a higher momentum:
\begin{eqnarray}
&&
\frac{\partial f_k(x,t)}{\partial t}
\!+\! v_k\frac{\partial f_k(x,t)}{\partial x}
\!+\! \frac{eE(x,t)}{\hbar}\frac{\partial f_k(x,t)}{\partial k}
=
- C_k[f](x,t).
\label{05}
\end{eqnarray}
here $v_k$ is the group velocity in the conduction band of the electrons,
and the field $E$ is the sum of the external nonconservative EMF, which does
{\em not} satisfy the Poisson Equation (\ref{02}) and a reactive
part that satisfies Equation (\ref{02}) and so is conservative.

Crucially, the right-hand side of Equation ({\ref{05}) consists of an Ansatz
encoding in $C$ the collisional effects that tend to restore the
distribution to its equilibrium Fermi-Dirac form as well as dissipating
the energy gained from the driving field
\cite{WW}.
Since the local carrier-number density (leading factor of two is for spin)
and its current are given by
\begin{equation*}
\rho(x,t) = 2\! \int \frac{dk}{2\pi} f_k(x,t) ~~{\rm and}~~ J(x,t)
= 2\! \int \frac{dk}{2\pi} v_k f_k(x,t),
\end{equation*}
conservation demands that, to recover Equation (\ref{01}), integration on
both sides of Equation (\ref{05}) requires
\begin{eqnarray}
\int \frac{dk}{2\pi} C_k[f](x,t) \equiv 0
\label{06}
\end{eqnarray}
unconditionally
\cite{note2}.
Although the Boltzmann collision integral can be quite complicated,
for practical use it may often be modelled more simply so long as
Equation (\ref{06}) is respected.
In a one-dimensional channel, conductance is defined as the magnitude
\[
G \equiv -e\frac{\partial J}{\partial V_{\rm sd}}
\]
where $V_{\rm sd}$ is the source-to-drain EMF over the channel
in the direction opposite to the flow $J$ defined as positive
(given electrons as the carriers of charge $-e$).
\begin{figure}
\centerline{
\hskip 5mm
\includegraphics[height=6.5 cm]{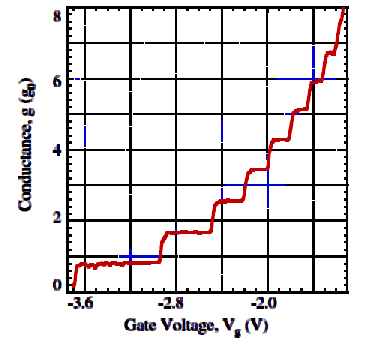}
}
\vskip 5mm
\caption{Landauer conductance measured in a QPC
\cite{1,dp2}, adapted from dePicciotto {\em et al}
\cite{dp2}.
Conductance is in units of the ideal step, $g_0 \approx 77.5\mu$S.
The experimental device was almost ideal, showing the low-field
conductance rising in its characteristic steps, increasing discretely
as each subband becomes newly occupied under the control of a gate voltage.
The carriers in each band thus appear to contribute independently.
}
\label{F3}
\end{figure}

The important feature to keep in mind for the transport problem
is that total charge neutrality for the QPC plus its screening
boundaries, of operational length $L$
as depicted in Figure \ref{F2}, means that the total carrier number
$N = \int^L_0 dx \sum_k f_k(x,t)$ is fixed by the compensating background.
It is independent of the current, This global constraint on the physics
is as important as the identity (\ref{06}) securing local conservation.

To introduce the experimental context we refer to the results shown
in Figure \ref{F3},
also due to dePicciotto and collaborators
\cite{1,dp2}.
In a quasi-one-dimensional (1D) quantum point contact, the restricted
geometry results in a series of discrete subbands offset by energy gaps
while, within each subband, carrier motion along the device remains free.
At low driving currents the occupancies of the bands, and their contribution
to device conductance, can be assumed to be uncoupled.

As the carrier density is increased via $V_g$, the capacitive gate voltage
controlling the chemical potential, there is successive occupation of higher
bands. Conductance increases in characteristic Landauer steps, each
contribution from the carriers in the newly populated subband acting fully
independently of its predecessors. Figure \ref{F3} illustrates this
weak-field driven regime. Other than a slight nonideal shortfall in
magnitude, the total device conductance increases stepwise as soon as the
gate-controlled carrier density accesses each successively higher subband.
There is no evidence of cross-talk or dynamical coupling among the discrete
populations. As the kinetic-theoretical approach does
\cite{WW},
standard single-particle accounts also fit this and many other comparable
measurements of QPC linear response.

\begin{figure}
\centerline{
\hskip 5mm
\includegraphics[height=6.75 cm]{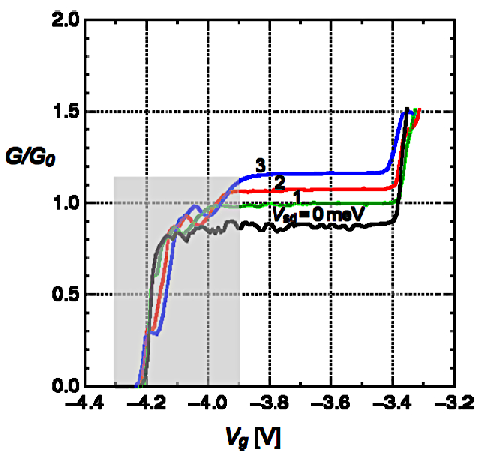}
}
\vskip 5mm
\caption{QPC conductance $G$ adapted from dePicciotto {\em et al}
\cite{1,dp2}
showing pronounced deviation of the Landauer steps from standard predictions. 
(Greyed-out band-threshold region covers the 0.7 anomaly
\cite{pepper},
a quite different effect at onset, not of interest here.)
Vertical scale for $G$ is again in Landauer units.
Nominally the gate voltage is set to restrict carrier density to the lowest
occupied subband while $V_{\rm sd}$, the driving EMF along the channel,
systematically increases. The QPC at low field ($V_{\rm sd} \!=\! 0) $ shows
near-ideal behaviour, as in Figure \ref{03}. At progressively higher EMF,
the step size of $G$ increases and indeed {\em exceeds} the Landauer limit.
This behaviour is not explicable by assuming carriers to move as single
particles, within uncoupled bands. What is needed is a fully conserving
kinetic description able to deal with dynamical interband cross-talk.
}
\label{F4}
\end{figure}

\subsection{Departure from linearity}

Figure \ref{F4}
\cite{1,dp2}
presents a strikingly different scenario at fields substantially above the
weak regime of Figure \ref{F3}.
In this experiment of dePicciotto and co-authors, the first interest was
to explore the ``0.7 anomaly'', manifesting at the ground-state band
threshold where $G$ first starts to rise. The 0.7 phenomenon has been
widely discussed, as evidenced not only by Refs. \cite{1,dp2} but also
in the representative Refs.
\cite{pepper,adam,DGqpc}
and in the manifold citations therein.

In our research, here reviewed, we address the entirely distinct nonlinear
response uncovered by the dePicciotto {\em et al} experiment. This novel
type of QPC behaviour has not been mentioned, let alone analysed
theoretically, anywhere in the 0.7 literature, other than in Refs.
\cite{1,dp2}
themselves.

At densities above the ground-state band threshold encompassing
the 0.7 anomaly (greyed out in Figure \ref{F4} as it does not enter
our discussion).
the graph presents a series of traces for $G$, again as a function of $V_g$
modulating the carrier density. The density regime is kept nominally within
the lowest band, below the threshold for accessing the next higher band.
The reported device had a working length of $2\mu$m; at a driving voltage
$V_{\rm sd} = 1$V the internal field is $E = 5{\rm kVcm}^{-1}$, well beyond
the weak regime.

In this circumstance the conductance might have been expected to stay
within the Landauer limit as seen at low field: $G \leq G_0$.
Evidently this is not so.

The Landauer limit is seen to be easily exceeded in Figure \ref{F4} as
$V_{\rm sd}$ is raised, despite the carrier density's apparently remaining
too low to access the next band and so induce an additional step in $G$.
Moreover, the enhancement depends on the EMF and is manifestly nonlinear.
Any linear-response account of the stepwise conductance would clearly be
insufficient to explain its increasing nonlinear enhancement.
It would be even less satisfactory to treat it as a collisionless
single-particle transmission model.

A conserving approach based on the quantum Boltzmann Equation is able, with
relatively simple assumptions, to reproduce the strongly nonequilibrium
behaviour of Figure \ref{F4}. The full formal description can be found in Ref.
\cite{GD2}.
Here we sketch the concept behind it in terms of a  form of
Equation (\ref{05}) in which two components, separated by a band gap,
are coupled by collisional cross-talk in terms of interband scattering.
This acts alongside the ever-present intraband scattering mechanisms.

The physical problem is one of active generation-recombination kinetics.
There are two populations: one in the ground-state  subband and
(a no longer trivial) one in the next higher band. They couple through
an interband scattering channel admitting particle exchange. Any
carrier loss out of the low band must be reflected by an equal gain
in the high band, and conversely.

Suppose $f_{1k}$ is the low-band carrier number distribution and
$f_{2k}$ the higher one. The system of Equations to solve is
\begin{eqnarray}
\frac{\partial f_{1k}}{\partial t}
\!+\! v_k\frac{\partial f_{1k}}{\partial x}
\!+\! \frac{eE}{\hbar}\frac{\partial f_{1k}}{\partial k}
&=&
- C_k[f_1] - {\cal C}_{k; 1 \to 2};
\cr
\cr
\frac{\partial f_{2k}}{\partial t}
\!+\! v_k\frac{\partial f_{2k}}{\partial x}
\!+\! \frac{eE}{\hbar}\frac{\partial f_{2k}}{\partial k}
&=&
- C_k[f_2] - {\cal C}_{k; 2 \to 1};
\label{07}
\end{eqnarray}
for simplicity we assume a uniform channel with uniform EMF field $E$
\cite{note4}.

As with Equation (\ref{06}) the intraband collision terms $C_k[f_i],i= 1,2$
on the right-hand side of Equation (\ref{07}) must each conserve separately
{\em provided} the interband terms ${\cal C}$ ere absent and the bands stay
blind to each other. To guarantee overall conservation, namely
\[
2\! \int \frac{dk}{2\pi}(f_{k1} + f_{k2}) = \rho = {\rm constant},
\]
one has to arrange the physical model so
\begin{eqnarray*}
\sum^2_{i=1}\! \int \frac{dk}{2\pi}
{\Bigl( C_k[f_i] \!+\! {\cal C}_{k; i \to (3 \!-\! i)} \Bigr)}
\equiv 0.
\label{08}
\end{eqnarray*}
These constraints are not enough to close the coupled problem. An additional
constitutive relation is needed and is provided by maximising the excess free
energy of the nonequilibrium system
\cite{GD2}.

The physical picture is that the free-motion acceleration by the EMF causes 
hot-electron energy to accumulate in the carriers until the dissipation
rate from collisions matches the input power and this excess in
free energy can rise no further, thus establishing steady state.
The scenario is equivalent to the assumption of the balance of dynamical
energy flows.

\begin{itemize}
\item[]
{{\em The inflow of external energy to the ``reservoir'' of hot-electron
free energy, available\\
for transfer to another medium, matches 
outflow to that medium, the ideal heat bath.}}
\end{itemize}

\begin{figure}
\centerline{
\hskip 5mm
\includegraphics[height=6.75 cm]{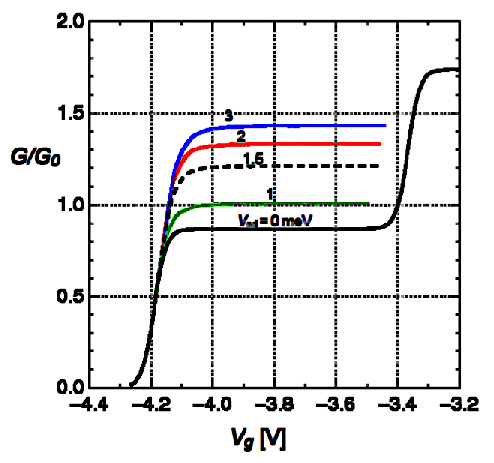}
}
\vskip 5mm
\caption{QPC conductance corresponding the experimental case, Figure \ref{F4},
computed within a conserving quantum Boltzmann model coupling two adjacent
channel subbands
\cite{GD2}.
At substantial driving fields, beyond the linear reguime, more carriers
from the lower band gain sufficient energy to cross the energy gap and
start to populate the higher band around its threshold, where $G$ is most
sensitive to small changes in population. The excited carriers enhance the
net conductance, exceeding the Landauer limit, before collisional
cross-coupling de-excites the promoted carriers to the lower band. In this
approach it is natural for enhancement of $G$ to be progressively greater
as the driving voltage $V_{\rm sd}$ ramps up.
}
\label{F5}
\end{figure}
\begin{figure}
\centerline{
\hskip 5mm
\includegraphics[height=6.75 cm]{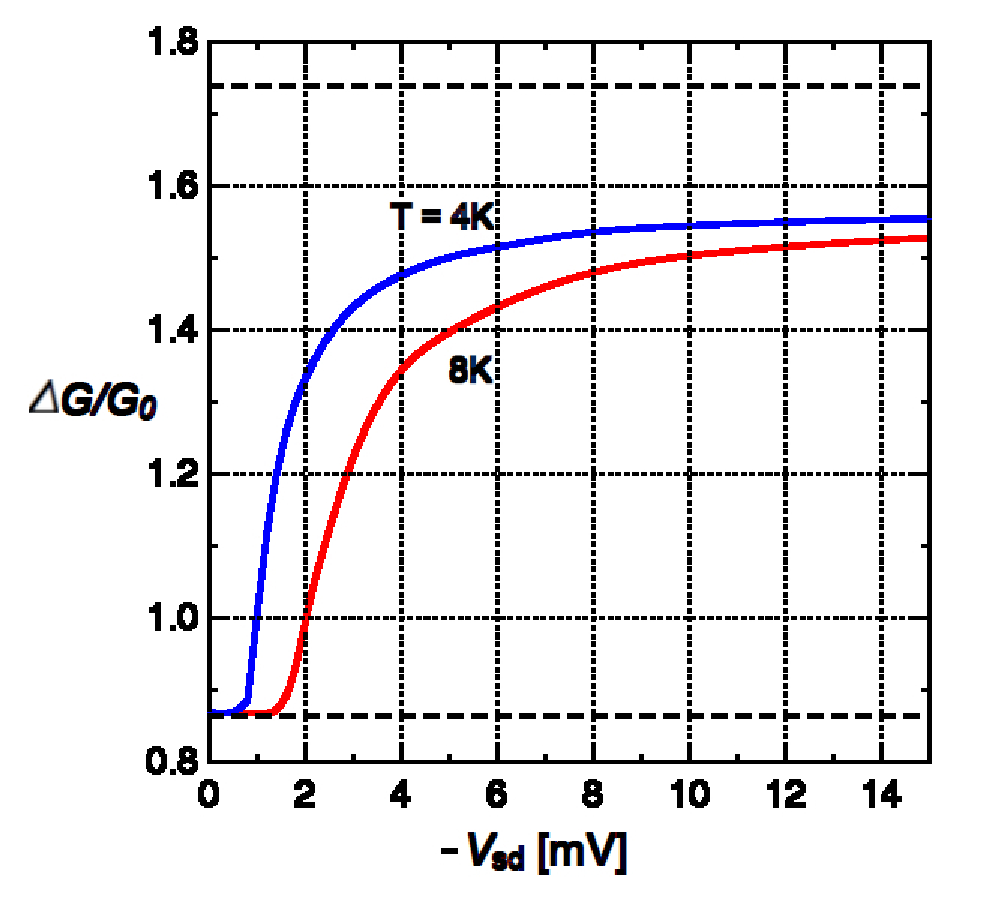}
}
\vskip 5mm
\caption{Threshold and saturation behaviour of excess conductance
$\Delta G$ versus source–drain voltage, at temperatures $T=4K$ and 8K
\cite{GD2}.
Lower dotted line is the base ground-state conductance step
upper is for the second sub-band, both replicating band data from Ref.
\cite{1}.
Two predictions for nonlinear QPC
response are made: (1) the threshold driving field for onset of the
enhancement scales approximately with $T$. (2) The saturation asymptote
at high fields is independent of temperature and undershoots the
contribution to $G$ from the next higher band.
}
\label{F6}
\end{figure}

Figure \ref{F5} displays our calculation
of nonlinear field enhancement of conductance in a QPC
\cite{GD2}.
The strong similarity to experiment is clear. While our theory
in its current form overemphasises nonlinear enhancement
with source-drain voltage, there remains room to improve the
quantitative accuracy. For instance, we have not implemented the
energy-dependent emission of optical phonons in our inelastic
relaxation rates. For an example of its role in hot-electron QPC
noise, see Ref.
\cite{gtd}.
In Figure \ref{F6}
we show additional properties of our nonequilibrium coupled-band kinetics,
from which two predictions follow: a common high-field saturation value
for the enhancement, and the temperature dependence of its onset and
growth rate with $V_{\rm sd}$.

The relative
simplicity of assumptions for the Boltzmann-Drude collision terms $C$ and
${\cal C}$ lead to a larger effect than seen in the experimental data
of Figure \ref{F4}. Refining the collision physics (for example, including
optical phonon emission) would improve a quantitative match.
Our priority has been to reproduce semi-quantitatively the salient aspects
of the nonlinear transport effect as measured. The quantum kinetic model
achieves this. and is open to refinement if new data were to emerge.
\begin{itemize}
\item
Allowing for transfer of highly excited carriers from the lower subband
to the higher produces an additional contribution to low-band conductance
not otherwise envisaged in weak-field particle transmission approaches.
\item
Competition between excitation and dynamical relaxation of promoted
carriers, assisted by stronger Pauli exclusion at higher densities,
leads to a robustly flat set of conductance plateaux.
\item
While conductance enhancement increases with increasing driving field and
readily exceeds the Landauer bound $G \leq G_0$, its rate of increase
tends to decline at larger EMF, pointing to saturation of the effect.
\end{itemize}

\section{Summary}

High-field conductance measurements of dePicciotto and co-workers
\cite{1,dp2}
were performed on exceedingly clean, almost ideal one-dimensional quantum
point contacts. A major finding of their investigation was the pronounced
nonlinearity of QPC conductance away from the more commonly explored linear
response regime.

The experiment we have reviewed challenges received theories of
quantised QPC conductance. It demands a more detailed kinetic theoretical
analysis that pays due attention, first and last, to particle conservation
at the level of its microscopic description. In this paper we have briefly
recalled a viable quantum Boltzmann theory of nonlinear QPC conductance
with reference to the above measurements.

The experimental results are already rather old, and have not
been revisited in more detail since. Even in a field moving rapidly to
sophisticated new types of excitations and their application to charge
transport at small scales, the problems at high driving fields have not
vanished. In fact they are likely to come into sharper focus than ever
as the scale of active structures becomes shorter yet.

In our view, a renewed set of investigations in the regime away from linear
response would stimulate more theoretical effort. That would provide a
rational grounding for new design tools in the nanoelectronic panorama
now emerging.


}
\end{singlespace}

\begin{thebibliography}{99}
\bibitem{note3}
The possibility of field lines inside the medium looping back
on themselves (that is, they are closed and multiply connected)
is not excluded by Eq. (\ref{03})
\cite{WW}.
\bibitem{note1}
In a two- or one-dimensional context, the chemical potential and thus the
charge for the system can be modulated, say capacitively, by an adjacent
gate. In that case the compensating charge can be physically offset from
the active structure; nevertheless bulk charge neutrality and
the physics of boundary screening apply.
\bibitem{note2}
On summing the left-hand side of Equation (\ref{05}), the term carrying
$\partial f_k/\partial k$ yields zero since the distribution $f_k$
must go to zero at large momenta (or else the sum diverges and so
would the density $\rho$).
\bibitem{note4}
The source of the driving field $E$ is external to the active
1D channel. Within the channel it is always so that
$\partial E/\partial x = 0$. This is not the case for a driving field
generated by mismatch of a continuous electrochemical potential (hence of
local carrier density). On the other hand, as in Ref.
\cite{GD3}
the role of screening from the macroscopic leads induces the carrier
density in a typical channel to be uniform, in which case only the
uniform, external, nonconservative EMF is the active driver.
\end{thebibliography}
\end{document}